\pdfoutput=1
\documentclass[sigconf]{acmart}
\usepackage{multirow}
\usepackage{graphicx}
\usepackage{subcaption}
\usepackage{listings}
\usepackage{listings}
\usepackage{xcolor}
\usepackage{longtable}
\usepackage{booktabs}
\usepackage{array}
\usepackage[ruled,vlined]{algorithm2e}

\AtBeginDocument{%
  \providecommand\BibTeX{{%
    \normalfont B\kern-0.5em{\scshape i\kern-0.25em b}\kern-0.8em\TeX}}}

\setcopyright{acmlicensed}
\copyrightyear{2018}
\acmYear{2018}
\acmDOI{XXXXXXX.XXXXXXX}

\begin{document}

% \title{HPCTransEval: A Benchmark of High Performance GPU-to-CPU Transpilation with Pre-trained Large Language Models}

% \title{Enhancing High Performance CUDA-to-CPU Transpilation with TVM-Driven Optimized Pretrained Large Language Models}

\title{HPCTransCompile: An AI Compiler Generated Dataset for High-Performance CUDA Transpilation and LLM Preliminary Exploration}

\author{Jiaqi Lv}
\authornote{Both authors contributed equally to this work.}
\affiliation{%
  \institution{Tongji University}
  \country{China}
}
\email{lvjiaqi@tongji.edu.cn}

\author{Xufeng He}
\authornotemark[1] % 引用上面的 \authornote
\affiliation{%
  \institution{Shanghai Artificial Intelligence Laboratory}
  \country{China}
}
\email{hexufeng@pjlab.org.cn}

\author{Yanchen Liu}
\affiliation{%
  \institution{Shanghai Artificial Intelligence Laboratory}
  \country{China}
}
\email{liuyanchen1@pjlab.org.cn}

\author{Xu Dai}
\authornote{Corresponding authors}
\affiliation{%
  \institution{Shanghai Artificial Intelligence Laboratory}
  \country{China}
}
\email{daixu@pjlab.org.cn}

\author{Aocheng Shen}
\affiliation{%
  \institution{Shanghai Artificial Intelligence Laboratory}
  \country{China}
}
\email{shenaocheng@pjlab.org.cn}

\author{Yinghao Li}
\affiliation{%
  \institution{Shanghai Artificial Intelligence Laboratory}
  \country{China}
}
\email{liyinghao@pjlab.org.cn}

\author{Jiachen Hao}
\affiliation{%
  \institution{Shanghai Artificial Intelligence Laboratory}
  \country{China}
}
\email{haojiachen@pjlab.org.cn}

\author{Jianrong Ding}
\affiliation{%
  \institution{Shanghai Artificial Intelligence Laboratory}
  \country{China}
}
\email{dingjianrong@pjlab.org.cn}

\author{Yang Hu}
\authornotemark[2]
% \authornote{Corresponding author} % 共同通讯作者
\affiliation{%
  \institution{Tsinghua University}
  \country{China}
}
\email{hu_yang@tsinghua.edu.cn}

\author{Shouyi Yin}
\affiliation{%
  \institution{Tsinghua University}
  \country{China}
}
\email{yinsy@tsinghua.edu.cn}

\begin{abstract}

% With the rapid development of deep learning, the model parameters have expanded exponentially. The demand for computational power has followed a similar trend according to the scaling law. 

The rapid growth of deep learning has driven exponential increases in model parameters and computational demands. NVIDIA GPUs and their CUDA-based software ecosystem provide robust support for parallel computing, significantly alleviating computational bottlenecks. Meanwhile, due to the cultivation of user programming habits and the high performance of GPUs, the CUDA ecosystem has established a dominant position in the field of parallel software. This dominance requires other hardware platforms to support CUDA-based software with performance portability. However, translating CUDA code to other platforms poses significant challenges due to differences in parallel programming paradigms and hardware architectures. Existing approaches rely on language extensions, domain-specific languages (DSLs), or compilers but face limitations in workload coverage and generalizability. Moreover, these methods often incur substantial development costs. Recently, LLMs have demonstrated extraordinary potential in various vertical domains, especially in code-related tasks. However, the performance of existing LLMs in CUDA transpilation, particularly for high-performance code, remains suboptimal. The main reason for this limitation lies in the lack of high-quality training datasets. To address these challenges, we propose a novel framework for generating high-performance CUDA and corresponding platform code pairs, leveraging AI compiler and automatic optimization technology. We further enhance the framework with a graph-based data augmentation method and introduce HPCTransEval, a benchmark for evaluating LLM performance on CUDA transpilation. We conduct experiments using CUDA-to-CPU transpilation as a case study on leading LLMs. The speedup ratio of the CPU operators has an average improvemnet of 43.8\%, highlighting the potential of LLMs to address compatibility challenges within the CUDA ecosystem. Our code is available at https://github.com/PJLAB-CHIP/HPCTransCompile.

\end{abstract}






\maketitle

\section{Introduction}

With the rapid growth of deep learning especially large language models (LLMs), the model parameters have expanded exponentially, and the demand for computing power shares the same trend according to the scaling law \cite{kaplan2020scalinglaws}. This trend drives the recent explosion of parallel software and hardware architectures. The most representative one is the NVIDIA GPU and its CUDA-based software ecosystem such as cuDNN, cuBLAS, etc. In fact, after years of continuous investment by NVIDIA in parallel computing, due to the cultivation of user programming habits and the high-efficiency performance GPUs, the CUDA ecosystem has already occupied a dominant position in the field of parallel software. This actuality requires other platforms such as CPUs, AMD GPUs, and TPUs \cite{jouppi2023tpu} to support CUDA-based software with performance portability.

Existing solutions for migrating CUDA software to other platforms primarily involve language extensions (e.g., OpenCL \cite{opencl}, SYCL \cite{sycl}) or specialized DSLs (e.g., Halide, TVM \cite{chen2018tvm}), often requiring manual intervention and specific library APIs \cite{yu2024enhancing}. Compiler-based approaches like Polygeist \cite{moses2023poly} automate translation but are limited by the completeness of their implementations and the range of CUDA workloads they support. These limitations highlight the unmet need for robust, generalizable tools capable of automatic and efficient CUDA transpilation.

The advent of large generative pre-trained language models has opened new possibilities for addressing this challenge. LLMs such as ChatGPT have demonstrated remarkable potential in code-related tasks, including code completion \cite{guo2023longcoder}, translation \cite{chen2018tree}, and repair \cite{jiang2021cure}. Tools like GitHub Copilot \cite{copilot} and Amazon CodeWhisperer \cite{amazon} demonstrate how LLMs can facilitate software development. However, the application of LLMs in translating CUDA to other platforms, particularly for high-performance code, remains underexplored. The core reason is the lack of high-quality training datasets relevant to the task. Although there are a large amount of high-performance CUDA and other platform codes available on open-source platforms, most of them lack descriptions and cannot be easily matched by functions. Moreover, most of the codes are highly integrated in various software, containing many definitions and types that only work within the specific software architecture. The codes collected from these programs are not conducive to enhancing the general knowledge of LLMs. Thus, it is vital to build a framework that can automatically generate CUDA code and corresponding platform code without being coupled to any specific software architecture.

To address these challenges, we propose a framework for generating high-performance CUDA and corresponding platform code pairs. This framework can leverage existing language extensions or specialized DSLs to generate datasets for LLM fine-tuning. Specifically, we use TVM as a demonstration tool to build this framework. TVM \cite{chen2018tvm} is a machine learning compiler that supports operator auto-optimization and code generation on various platforms including GPUs, CPUs, and other ASICs. Facilitated by the TVM auto-scheduler, which can produce efficient code that is competitive with human-optimized libraries, high-performance CUDA and other platform code pairs can be automatically generated by TVM. However, directly using TVM still presents challenges. One challenge is that the common operators supported in the TVM operator inventory (TOPI) limit the diversity of computations. Another challenge is that the generated code includes dependencies related to the TVM library. To tackle these challenges, we propose a graph-based data enhancement method to construct a representative and diverse dataset. We also modify the TVM framework to ensure that the generated code does not have specific dependencies.

Meanwhile, the CPU remains the most common and versatile computing platform in various applications. Re-engineering CUDA software for CPU platforms entails more substantial costs and risks, especially when systems must continue to evolve and adapt. Beyond practical necessity, the translation of CUDA to CPU code represents a broader academic challenge, exemplifying the need to bridge disparate parallel programming paradigms \cite{laso2024exploring}, namely CUDA's single-program multiple data (SPMD) model and the fork-join model used on CPUs. Therefore, we use the CUDA-to-CPU translation task as an entry point to systematically evaluate the potential of LLMs to address compatibility challenges within the CUDA ecosystem.

In addition, we carefully select representative operators to create HPCTransEval. This is a benchmark for high-performance CUDA transpilation. HPCTransEval includes 100 standalone operators, 100 computation graphs built from these operators, and 10 building blocks for deep learning models. The benchmark is independent of specific software architecture and can effectively evaluate LLMs' performance on CUDA transpilation tasks.

Experiments are conducted on three leading LLMs: Qwen2.5-Coder, DeepSeek-Coder-V2, and OpenCoder. These evaluations include analyses of overall and granular performance, the impact of dataset size, and insights into their respective strengths and limitations. The results highlight key challenges in high-performance CUDA transpilation, emphasizing the potential of leveraging automated code generation to fine-tune LLMs for improved compatibility and optimized performance.

In summary, the main contributions of this paper are as follows.

\begin{itemize}
\item We propose a framework for generating high-performance code pairs of GPU and corresponding platform. This framework can leverage language extensions or specialized DSLs for automated code generation to create datasets for LLM fine-tuning, improving their performance in CUDA transpilation task.
\item We introduce HPCTransEval, a high-performance benchmark for CUDA transpilation. This benchmark is architecture-agnostic and can effectively evaluate LLM performance on this task.
\item We conduct experiments using CUDA-to-CPU transpilation as a case study. The results on HPCTransEval and another benchmark demonstrate that our framework significantly enhances LLMs performance in CUDA transpilation, particularly for high-performance code.
\end{itemize}

\section{Background}

In this section, we will first introduce the state-of-the-art large language models in the field of code. Next, we will discuss TVM, the demonstration tool used to automatically generate CUDA and corresponding CPU C code pairs.

\subsection{Large Language Models for Code Task}

Large language models (LLMs) have demonstrated remarkable capabilities in code-related tasks, including code completion, synthesis, and transpilation. Among these, Qwen2.5-Coder \cite{hui2024qwen2}, DeepSeek-Coder-V2 \cite{zhu2024deepseek}, and OpenCoder \cite{huang2024opencoder} represent significant advances. Qwen2.5-Coder is an open-source model tailored for multilingual and multi-platform code generation and understanding tasks. Building upon the strengths of its predecessor, Qwen-Coder, it incorporates enhanced tokenization strategies and advanced architectural features such as grouped query attention and expanded context windows. Qwen2.5-Coder's training process emphasizes fill-in-the-middle and other code-specific objectives, leveraging extensive datasets that capture diverse programming paradigms and patterns. DeepSeek-Coder-V2 is a Mixture-of-Experts (MoE) \cite{shazeer2017outrageously} model designed for high-performance coding tasks. It substantially improves upon DeepSeek-V2 by extending its support to 338 programming languages, up from 86, and increasing the context window length from 16,000 tokens to an impressive 128,000 tokens. This expansive context length allows DeepSeek-Coder-V2 to excel in tasks requiring extensive contextual understanding, such as code transpilation. Additionally, it enhances coding and mathematical reasoning capabilities while maintaining competitive performance in general language tasks. OpenCoder, another notable contribution, focuses on efficient code translation across different languages and paradigms. By leveraging state-of-the-art transformer architectures and pretraining techniques, OpenCoder achieves high performance in cross-language code understanding and transformation tasks. It introduces innovative mechanisms for semantic alignment and syntax preservation, enabling accurate and context-aware code transpilation.

\subsection{Tensor Virtual Machine (TVM)}

TVM \cite{chen2018tvm} is an open-source machine learning compiler framework for CPUs, GPUs, and other ASICs which is widely used both in academia and industry. The most appealing aspect of TVM is the automatic optimization of operators. TVM utilizes a scheduling language that can describe loop optimization primitives such as tiling, unrolling, parallelization, etc motivated by Halide \cite{ragan2013halide} compiler. The separation of scheduling language and computation abstraction makes TVM suitable for both manual optimization and automatic search. TVM supports both template-guided search (AutoTVM) and template-free search (auto-scheduler) to find the best scheduling primitives with related configurations. The auto-scheduler known as Ansor \cite{zheng2020ansor} can automatically generate a large search space with comprehensive coverage of optimizations and is able to find high-performance programs that can even compete with human experts. In this work, we utilize an auto-scheduler to search best scheduling configurations both for various operators in TVM operator inventory (TOPI) and generated computation graph workloads.

\section{Framework Implementation}

In this section, we will introduce the details of the proposed framework. The entire framework consists of six components: operator selection, computation graph construction, useless dependency removal, graph labeling, dataset generation, and evaluation process. The framework is illustrated in Figure \ref{fig:evaluate_process}.

\begin{figure*}
    \centering
    \includegraphics[width=0.9\linewidth]{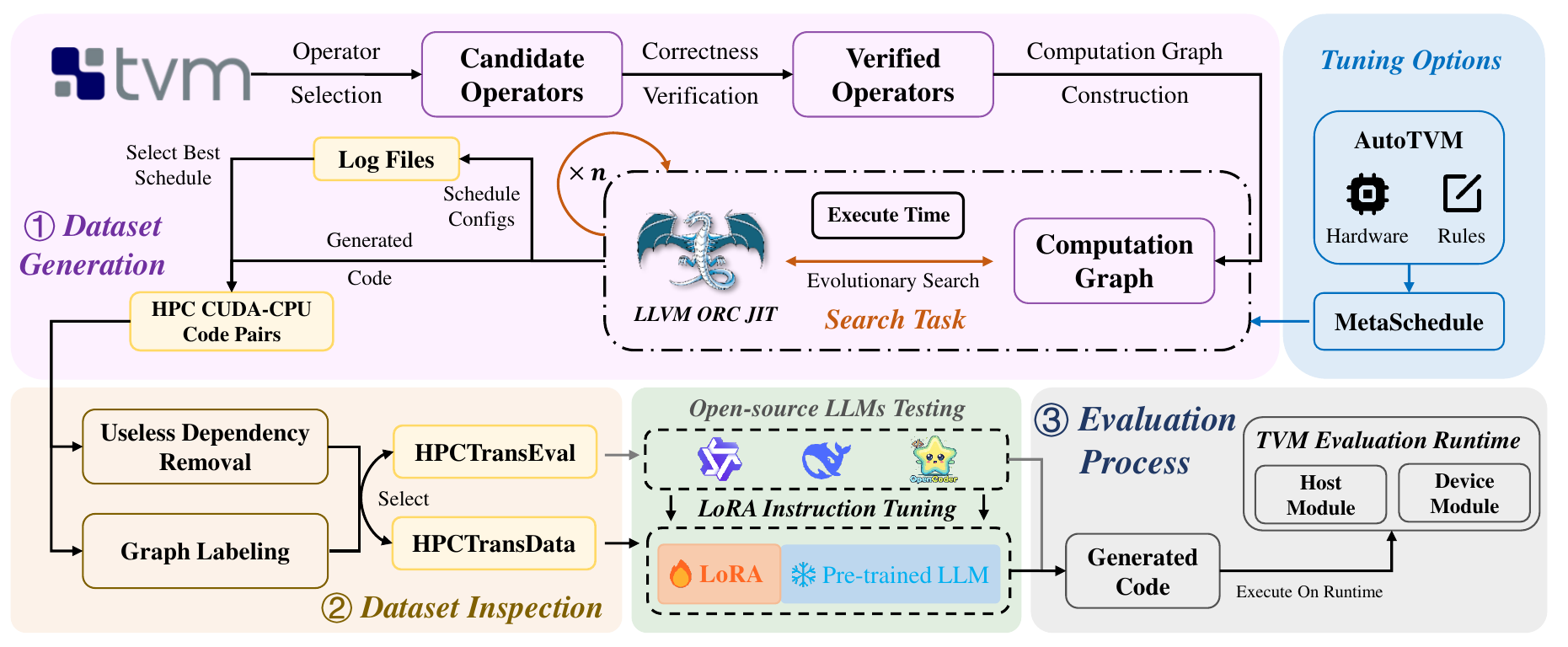} % 设置宽度并指定图片路径
    \caption{Implementation of the proposed framework.} % 标题
    \label{fig:evaluate_process} % 标签（用于交叉引用）
\end{figure*}

\subsection{Operator Selection}

\begin{table*}[h]
\centering
\caption{Operator Types and Their Descriptions}
\label{table:op_type_and_description}
\begin{tabular}{p{3cm}|p{5cm}|p{5cm}}
\hline
\textbf{Operator Type} & \textbf{Description} & \textbf{Examples} \\
\hline
$Elementwise$ & Performs operations independently on each element of the input data. & add, sub, mul, div, sqrt, sin, cos, relu and etc. \\
\hline
$Reduction$ & Aggregates elements along specified axes to
obtain a lower-dimensional output. & sum, mean, argmax, argmin, softmax and etc. \\
\hline
$LayoutTransform$ & Reorganizes data layout without altering values. & reshape, transpose and etc. \\
\hline
$LogicIntensive$ & Involves complex conditional or logical operations dominating computation. & topk, sort, grid sample etc. \\
\hline
$ComputeIntensive$ & Requires heavy numerical computations, often with high arithmetic intensity. & matmul, conv, einsum, dft etc. \\
\hline
\end{tabular}
\end{table*}

To make the generated dataset usable and diverse, we first filter the operators currently supported by TVM. We select operators by following steps: (1) collect the information of candidate operators from the TVM official documentation, including the operator name, call inference, function description, and parameter settings. (2) verify the correctness of each independent operator. If the operator can generate the corresponding GPU and CPU code through the TVM $CodeGen$ module, we will add the operator to the available list. If the operator is not supported by the $CodeGen$ module, we will manually implement it. (3) divide the operators into five categories according to the computation patterns: Elementwise, Reduction, Layout Transform, Logic Intensive, and Compute Intensive. The detailed description of each category is shown in Table \ref{table:op_type_and_description}.

\subsection{Computation Graph Construction}

% \begin{figure}
%     \centering
%     \includegraphics[width=0.95\linewidth]{computation_graph_construction.pdf} % 设置宽度并指定图片路径
%     \caption{The construction process of computation graph.} % 标题
%     \label{fig:evaluate_process} % 标签（用于交叉引用）
% \end{figure}

\begin{lstlisting}[language=C++, caption={CUDA Device Code.},label={lst:cuda_code}, breaklines=true]
extern "C" __global__ void __launch_bounds__(7) default_function_kernel(float* __restrict__ compute, float* __restrict__ data) {
    compute[((((int)blockIdx.x) * 7) + ((int)threadIdx.x))] = __cosf(data[((((int)blockIdx.x) * 7) + ((int)threadIdx.x))]);
}
\end{lstlisting}

\begin{lstlisting}[language=C++, caption={CPU C Code with TVM dependency.},label={lst:cpu_with_tvm}, breaklines=true]
TVM_DLL int32_t default_function(void* args, int32_t* arg_type_ids, int32_t num_args, void* out_ret_value, int32_t* out_ret_tcode, void* resource_handle) {
    void* compute = (((TVMValue*)args)[1].v_handle);
    void* compute_1 = (((DLTensor*)compute)[0].data);
    // Computation Loop
    for (int32_t i1 =0; i1 <36; ++i1) {
        for (int32_t i2 =0; i2 <9; ++i2) {
            int32_t cse_var_1 = ((i1  *9) + i2);
            ((float*)compute_1)[cse_var_1] =cosf(((float*)tarray_1)[cse_var_1]);
        }
    }
    return 0;
}
\end{lstlisting}

\begin{algorithm}
\SetAlgoLined
\KwIn{Set of candidate operators $\mathcal{O}$, Initial computation graph $\mathcal{G}_{init}$, Maximum number of nodes $N_{max}$, Maximum graph depth $D_{max}$, The probability of an operator being selected $P_{op}$.}
\KwOut{Final diverse set of computation graphs $\mathcal{G}_{final}$.}
Initialize the graph set $\mathcal{G} = \{\mathcal{G}_{init}\}$\;
\ForEach{$G \in \mathcal{G}$}{
    \Repeat{$|G| > N_{max}$ or $Depth(G) > D_{max}$}{
        \tcp{Candidate Node Expansion}
        Select expansion nodes $\mathcal{N}_{expand} \subseteq Nodes(G)$\;
        \ForEach{$n \in \mathcal{N}_{expand}$}{
            Heuristically select $op \in \mathcal{O}$ which satisfies $Usage(op) \le  P_{op} \times |\mathcal{O}|$\;
            Generate a new node $n_{new}$ associated with $op$ and insert it into $G$\;
        }
        \tcp{Introduce Topological Diversity}
        Randomly select an action from the following:\\
        (1) Add random connection: $P(n_i \to n_j) \propto \frac{InDegree(n_j)}{Degree(G)}$\;
        (2) Add branch structures: $n_{branch} \to \{n_{new_1},n_{new_2},\dots\}$\;
        (3) Merge paths: $n_{merge} \to n_{target}, \forall n_{merge} \in Path(G)$\;
        \tcp{Ensure data flow validity}
        $\forall \text{path}_i \in \text{Path}(G), \text{FlowValid}(\text{path}_i)$\;
    }
    $\mathcal{G}_{final} \gets \mathcal{G}_{final} \cup G$\;
}
\caption{Computation Graph Construction Algorithm}
\label{alg:graph_construction}
\end{algorithm}

After the operator selection, we construct the computation graph using the operators in the available list. We aim to provide a comprehensive evaluation of the high-performance GPU-to-CPU transpilation task. This requires a diverse dataset that encompasses different types and complexities of computations. Therefore, when constructing the computation graph, we consider both the types of operators included in the graph and the overall computational complexity. Specifically, we repeat to select an operator and extend it onto the graph. Meanwhile, we employ methods such as random connections, branch expansion, and branch merging to enhance the complexity of the computational graph. The nodes in the computation graph represent computation types, while the edges represent the flow of data.

After each expansion of the computation graph, we validate the legality of the computation flow. First, the computation graph must be a directed acyclic graph (DAG). Next, we verify the consistency of node operations. Each node's operation type must match the data types and shapes of its inputs and outputs. Additionally, all computational nodes must have their inputs derived from either input nodes or other computational nodes to avoid "dangling" nodes. If the current expansion introduces errors, the expansion will be rolled back and a new random expansion will be performed until the final constraints are satisfied. A complete computation graph construction process is illustrated in Algorithm \ref{alg:graph_construction}.

\subsection{Useless Dependency Removal}

% \begin{lstlisting}[language=C++, caption={CUDA Device Code.},label={lst:cuda_code}, breaklines=true]
% extern "C" __global__ void __launch_bounds__(7) default_function_kernel(float* __restrict__ compute, float* __restrict__ data) {
%     compute[((((int)blockIdx.x) * 7) + ((int)threadIdx.x))] = __cosf(data[((((int)blockIdx.x) * 7) + ((int)threadIdx.x))]);
% }
% \end{lstlisting}

After the previous computation graph construction step, the dataset has already contained various computation patterns. However, there is still an issue that needs to be addressed: the CPU C code produced by the TVM code generator contains a large number of dependencies defined by the TVM library itself which may influence model fine-tuning and evaluation performance. More specifically, the original TVM CPU platform compilation passes mix host code and device code in a single module since both codes ultimately run on the CPU. However, this also results in the generated code containing the specific library dependency codes. Therefore, we modify several related compilation passes for the CPU platform to separate the host and device into two modules while satisfying the calling conventions for kernel launching. In the code generation phase, the host and device modules call their corresponding code generators. Additionally, for the device module code generator, we avoid generating codes containing TVM library-specific header files including, macros definitions and TVM API calls to produce clean CPU C device code. The CUDA code, along with the TVM-based CPU code and the TVM-free CPU code, is shown in Listing \ref{lst:cuda_code}, \ref{lst:cpu_with_tvm}, and \ref{lst:cpu_without_tvm}, respectively.

\subsection{Graph Labeling}

For large language models, the prompt is a crucial factor that determines the performance \cite{zhou2022large} of LLMs. In the domain of code transpilation, the task specification can significantly effect  the quality of the generated code. Appropriate prompts can guide the model to generate code that better aligns with the task requirements and contextual constraints, while unsuitable prompts may result in code with logical errors, compilation issues or unexpected behavior. In the high-performance GPU-to-CPU transpilation task, hardware information such as processor architecture, available core and thread counts and cache capacities can also influence the generated code. Therefore, we record the utilized hardware information as part of the annotation for transpilation task. Additionally, we recruit professional high-performance engineers to provide computational flow descriptions and hardware-aware optimization strategies as function-level annotations, which serve to explain the internal logic and provide viable optimization paths.

\subsection{Dataset Generation}

We use TVM to represent the computation flow of the graph and encapsulate it into a function decorated with $auto scheduler$ module. Then, we combine this function with the target (LLVM on CPU, CUDA on GPU) to form a Search Task, and specify the tuning methods. During the search, we set the number of measurement trials to 200. The CPU we use is the Intel(R) Xeon(R) Gold 6348 CPU @ 2.60GHz with 112 cores.

TVM will automatically search for the optimal schedule and save the searched configuration in a log file. When the search times reaches the set threshold or exceeds the maximum search time, TVM will stop the search task and use the saved optimal configuration to generate code for the corresponding hardware platform. If no valid schedule is found within the specified time, the search task for the next computation graph will proceed.

\begin{lstlisting}[language=C++, caption={CPU C Code without TVM dependency.},label={lst:cpu_without_tvm}, breaklines=true]
void default_function_kernel(float* compute, float* data) {
    #pragma omp parallel for
    // Computation Loop
    for (int32_t i1  =0; i1  <36; ++i1 ) {
        for (int32_t i2 =0; i2 <9; ++i2) {
            compute[i1] =cosf(data[i1*9+i2]);
        }
    }
}
\end{lstlisting}

\subsection{Evaluation Process}

To evaluate the correctness of the generated CPU code, we need an evaluation runtime to enable automated code correctness testing that supports loading, compiling, and executing source codes. However, the original TVM CPU C source runtime module does not support these features mainly due to two reasons: 1) As mentioned before, the compilation process of the TVM CPU platform mixes host and device code in a single module, resulting in a tightly coupled generated code that makes it difficult to replace the kernel functions with externally provided codes; 2) the CPU C source runtime module only supports code export and does not support execution. Therefore, we extend the original TVM runtime with a new runtime called source code evaluation runtime, which supports all the features needed for automatic testing.

The detailed implementation of source code evaluation runtime is shown in Figure \ref{fig:evaluate_process}. This runtime provides source code loading, compilation, and execution APIs for users to test externally provided codes while implementing necessary functions to integrate with the TVM runtime framework. It is worth mentioning that the generated dataset also contains exported host and device runtime modules with additional meta information corresponding to CUDA and CPU C code pairs for evaluation. During the evaluation process, for each test case, the related exported host and device runtime modules with meta information are loaded first. Secondly, the CPU C code is loaded to the device module to be compiled with auto-generated glue codes to match calling conventions. The host module is then executed via LLVM ORC JIT \cite{orcjit} which calls compiled device kernel functions to produce the final results. During the loading, compiling, and execution phases, the runtime can return the corresponding errors once compilation or execution fails.

\begin{figure}
    \centering
    \includegraphics[width=1.0\linewidth]{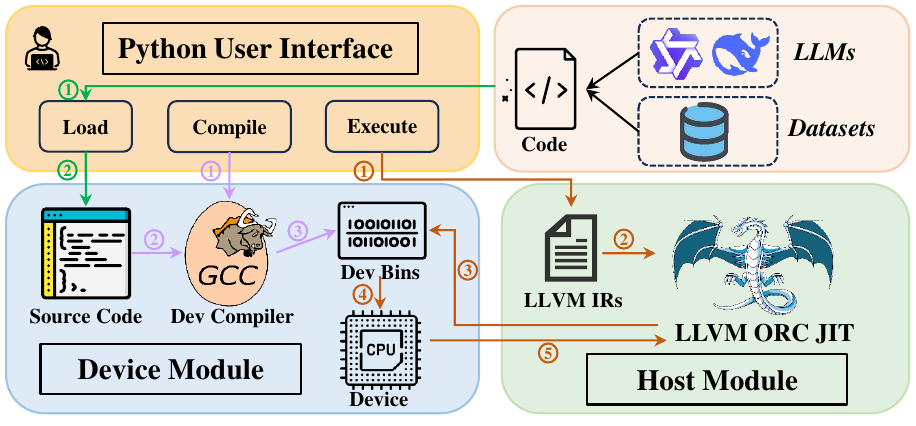} % 设置宽度并指定图片路径
    \caption{Source Code Evaluation Runtime.} % 标题
    \label{fig:evaluate_process} % 标签（用于交叉引用）
\end{figure}

\section{Experimental Setup}

In this section, we will introduce the dataset and model used in the experiments. Additionally, we provide a detailed description of the training details, including parameter configurations and training epochs during the model fine-tuning process.

\subsection{Datasets And Models}

We leverage our proposed framework to generate a dataset, HPCTransData, comprising 20k CUDA-CPU code pairs. This dataset serves as the foundation for fine-tuning the selected models. In our experiments, we select state-of-the-art LLMs that support multilingual code generation to ensure a comprehensive evaluation. Specifically, we select three open-source models, Qwen2.5-Coder, DeepSeek-Coder-V2-Lite, and OpenCoder for our study. Subsequently, they will be referred to as $\mathcal{M}_1$, $\mathcal{M}_2$, and $\mathcal{M}_3$, respectively. Experiments are conducted on two benchmarks, HPCTransEval, and $KernelBench_c$. HPCTransEval consists of manually curated and supplemented code generated by TVM. $KernelBench_c$ is constructed on KernelBench \cite{ouyang2025kernelbench} and introduced to specifically assess the generalization capabilities of our framework. By utilizing $KernelBench_c$, we aim to evaluate how well the framework adapts to diverse scenarios and novel code patterns, providing a more rigorous validation of its versatility and robustness in CUDA-to-CPU transpilation tasks.

\textit{HPCTransEval}: HPCTransEval is a carefully curated set of representative primitive operators and fused computation graphs selected from CUDA-CPU code pairs generated based on TVM. Additionally, HPCTransEval incorporates widely used model modules to enhance the comprehensiveness of the benchmark. Specifically, HPCTransEval includes 100 primitive operators, 100 fused computation graphs, and 10 complex models. The code in this benchmark is free from dependencies on any specific framework.

\textit{$KernelBench_c$}: The KernelBench benchmark is a set of 250 neural network tasks defined as PyTorch modules. The tasks are split into three categories denoted as levels 1, 2, and 3. Level 1 tasks are common ML primitives, such as softmax or various matmuls. Level 2 tasks are few-step fusions of those primitives, such as network layer activations followed by a norm. Level 3 tasks represent full network architectures (e.g., a ResNet). We extend KernelBench by manually writing CUDA code corresponding to PyTorch modules, forming $KernelBench_c$.

\subsection{Implementation Details}

\textit{Training Hyperparameters}: We implement LLMs fine-tuning using the HuggingFace Transformers libirary v4.45.0 \cite{wolf2020transformers}. We initialize the tokenizer with the pre-trained tokenizer from the original model, which is further trained on our specific training dataset. The training is conducted using the AdamW optimizer \cite{loshchilov2017decoupled} and a batch size of 32 for 5 epochs. The low-rank adaptation \cite{hu2021lora} (LORA) is adopted in the fine-tuning process with a rank of 8, an alpha value of 16, and a dropout rate of 0.05, specifically targeting the key projection layers ($q_{proj}, k_{proj}, v_{proj}$). Moreover, The gradient accumulation steps are set to 2, and the warmup ratio of 0.3 and linear learning rate scheduler are applied. All experiments are run on a single node with eight NVIDIA A100 SXM4 GPUs, each with 80GB of memory. To speed up the training process, mixed-precision training was enabled. The trainable parameters account for approximately 10\% of the original model.

\textit{Evaluation Metric}: To evaluate the performance of different LLMs before and after fine-tuning, we utilize three key metrics: \textbf{Compile Pass}, \textbf{Execute Pass}, and \textbf{Speedup Ratio}. The Compile Pass metric assesses whether the translated CPU code can be compiled successfully without errors, ensuring the correctness of syntax and dependencies. The Execute Pass metric verifies that the translated CPU code executes correctly, producing expected results within a reasonable error margin. Finally, the Speedup Ratio measures the performance improvement, calculated by comparing the execution time of the translated CPU code to the original code supported by PyTorch or TVM.

\section{Experimental Results}

In this section, we first present the overall performance of three state-of-the-art models on $HPCTransEval$ and $KernelBench_c$. Next, we focus on analyzing the performance differences of the code before and after fine-tuning, providing specific examples for illustration. Finally, we conduct a further analysis of the impact of dataset size on performance improvement.

\subsection{Overall Effectiveness}

% Model List
% Qwen2.5-Coder-14B | DeepSeek-Coder-V2-Lite | OpenCoder-8B

\begin{table*}[t]
\caption{Performance Comparison Across Three Metrics Before and After Fine-Tuning in $KernelBench_c$. The Last Column Highlights the Improvements.}
\centering
\begin{tabular}{c|c|lll|lll|lll}
\toprule
\multirow{2}{*}{Metric} & \multirow{2}{*}{Model} & \multicolumn{3}{c|}{Level 1 $[Basic Ops]$} & \multicolumn{3}{c|}{Level 2 $[Fused Ops]$} & \multicolumn{3}{c}{Level 3 $[Complex Models]$} \\ 
\cmidrule(lr){3-5} \cmidrule(lr){6-8} \cmidrule(lr){9-11}
                        &                       & Before   & After   & Improvement   & Before   & After   & Improvement   & Before   & After   & Improvement   \\ 
\midrule
\multirow{3}{*}{Compile Pass}      
                        & $\mathcal{M}_1$           & 0.39     & 0.60     & +0.21 (↑53.8\%)  & 0.42     & 0.51     & +0.09 (↑21.4\%)  & 0.32     & 0.58     & +0.26 (↑81.3\%) \\
                        & $\mathcal{M}_2$           & 0.44     & 0.47     & +0.03 (↑6.8\%)   & 0.34     & 0.40     & +0.06 (↑17.6\%)  & 0.54     & 0.58     & +0.04 (↑7.4\%)  \\
                        & $\mathcal{M}_3$           & 0.22     & 0.26     & +0.04 (↑18.2\%)  & 0.22     & 0.33     & +0.11 (↑50.0\%)  & 0.26     & 0.40     & +0.14 (↑53.8\%) \\
\midrule
\multirow{3}{*}{Execute Pass}  
                        & $\mathcal{M}_1$           & 0.28     & 0.51     & +0.23 (↑82.1\%)  & 0.35     & 0.39     & +0.04 (↑11.4\%)  & 0.22     & 0.40     & +0.18 (↑81.8\%) \\
                        & $\mathcal{M}_2$           & 0.14     & 0.17     & +0.03 (↑21.4\%)  & 0.15     & 0.18     & +0.03 (↑20.0\%)  & 0.26     & 0.38     & +0.12 (↑46.2\%) \\
                        & $\mathcal{M}_3$           & 0.06     & 0.08     & +0.02 (↑33.3\%)  & 0.12     & 0.13     & +0.01 (↑8.3\%)   & 0.18     & 0.30     & +0.12 (↑66.7\%) \\
\midrule
\multirow{3}{*}{Speedup Ratio}     
                        & $\mathcal{M}_1$           & 2.90     & 6.75    & +3.85 (↑132.8\%) & 3.45     & 8.20     & +4.75 (↑137.7\%) & 1.46     & 1.89     & +0.43 (↑29.5\%) \\
                        & $\mathcal{M}_2$           & 1.94     & 2.06     & +0.12 (↑6.2\%)   & 1.09     & 1.15     & +0.06 (↑5.5\%)   & 1.21     & 1.62     & +0.41 (↑33.9\%) \\
                        & $\mathcal{M}_3$           & 0.82     & 1.13     & +0.31 (↑37.8\%)  & 0.84     & 0.91     & +0.07 (↑8.3\%)   & 0.95     & 0.97     & +0.02 (↑2.1\%)  \\
\bottomrule
\end{tabular}
\label{tab:fine_tune_comparison_kernelbench_c}
\end{table*}

\begin{table*}[t]
\caption{Performance Comparison Across Three Metrics Before and After Fine-Tuning in $HPCTransEval$. CP, EP, and SR represent Compile Pass, Execute Pass, and Speedup Ratio, respectively.}
\centering
\begin{tabular}{c|c|ccc|ccc|ccc}
\toprule
\multirow{2}{*}{Level} & \multirow{2}{*}{Tuning} & \multicolumn{3}{c|}{$\mathcal{M}_1$} & \multicolumn{3}{c|}{$\mathcal{M}_2$} & \multicolumn{3}{c}{$\mathcal{M}_3$} \\ 
\cmidrule(lr){3-5} \cmidrule(lr){6-8} \cmidrule(lr){9-11}
                        &                       & CP   & EP   & SR   & CP   & EP   & SR   & CP   & EP   & SR   \\ 
\midrule
\multirow{2}{*}{Level 1 $[Basic Ops]$}      
                        & $Before$           & 0.54     & 0.30     & 0.83  & 0.45     & 0.12     & 0.87  & 0.20     & 0.14     & 0.76  \\
                        & $After$               & 0.87     & 0.62     & 2.35   & 0.74     & 0.58     & 2.04  & 0.82     & 0.51     & 1.74  \\

\midrule
\multirow{2}{*}{Level 2 $[Fused Ops]$}  
                        & $Before$               & 0.43     & 0.24     & 1.04   & 0.36     & 0.20     & 0.97   & 0.27     & 0.19     & 0.85 \\
                        & $After$               & 0.74     & 0.63     & 1.44  & 0.65     & 0.62     & 1.68  & 0.57     & 0.44     & 1.23 \\
\bottomrule
\end{tabular}
\label{tab:fine_tune_comparison_hpctranseval}
\end{table*}

As shown in Table \ref{tab:fine_tune_comparison_kernelbench_c}, for $KernelBench_c$, the fine-tuning results reveal that the injection of domain-specific knowledge yields the greatest returns on simpler computational kernels while still benefiting more complex models. For $\mathcal{M}_1$, we observe a remarkable 82.1\% uplift in Execute Pass on basic operations (Level 1) and a 132.8\% jump in the Speedup Ratio. This evidences that targeted tuning dramatically accelerates code emit and kernel execution for low-level primitives. Even fused operations (Level 2) see a 50\% increase in Compile Pass and a 137.7\% rise in the Speedup Ratio for $\mathcal{M}_1$, underscoring the gains from embedding domain knowledge about operator fusion. Although $\mathcal{M}_2$ and $\mathcal{M}_3$ show smaller overall gains compared to $\mathcal{M}_1$, reflecting inherent differences in their base architectures. Substantial improvements after fine-tuning confirm that injecting kernel-level domain knowledge effectively enhances the ability of LLMs to translate code into high-performance implementations.

For $HPCTransEval$, before fine-tuning, the Speedup Ratios are mostly below 1, indicating inefficiencies in generated code. As shown in Table \ref{tab:fine_tune_comparison_hpctranseval}, fine-tuning significantly improved all metrics, especially the Speedup Ratios, which increased to more than 1 for both Basic Ops (Level 1, +183\%) and Fused Ops (Level 2, +38\%). This demonstrates enhanced code correctness and execution efficiency, although Fused Operators showed smaller improvements due to complexity. Additionally, due to the overly complex implementation of model building blocks in HPCTransEval, none of the operators could run correctly. This highlights the limitations of LLMs in handling more complex code translation tasks. In contrast, many tasks of $KernelBnech_c$ at level 3 are successfully completed, as these translations mainly involve direct mapping to Torch API calls. Thus, HPCTransEval, with its focus on kernel function translation, serves as a more valuable reference for advancing research on more granular and complex code translation scenarios.

\subsection{Performance Optimization Analysis}

\begin{figure*}[htbp]
    \centering
    \begin{minipage}{0.47\textwidth}
        \centering
        \includegraphics[width=\linewidth]{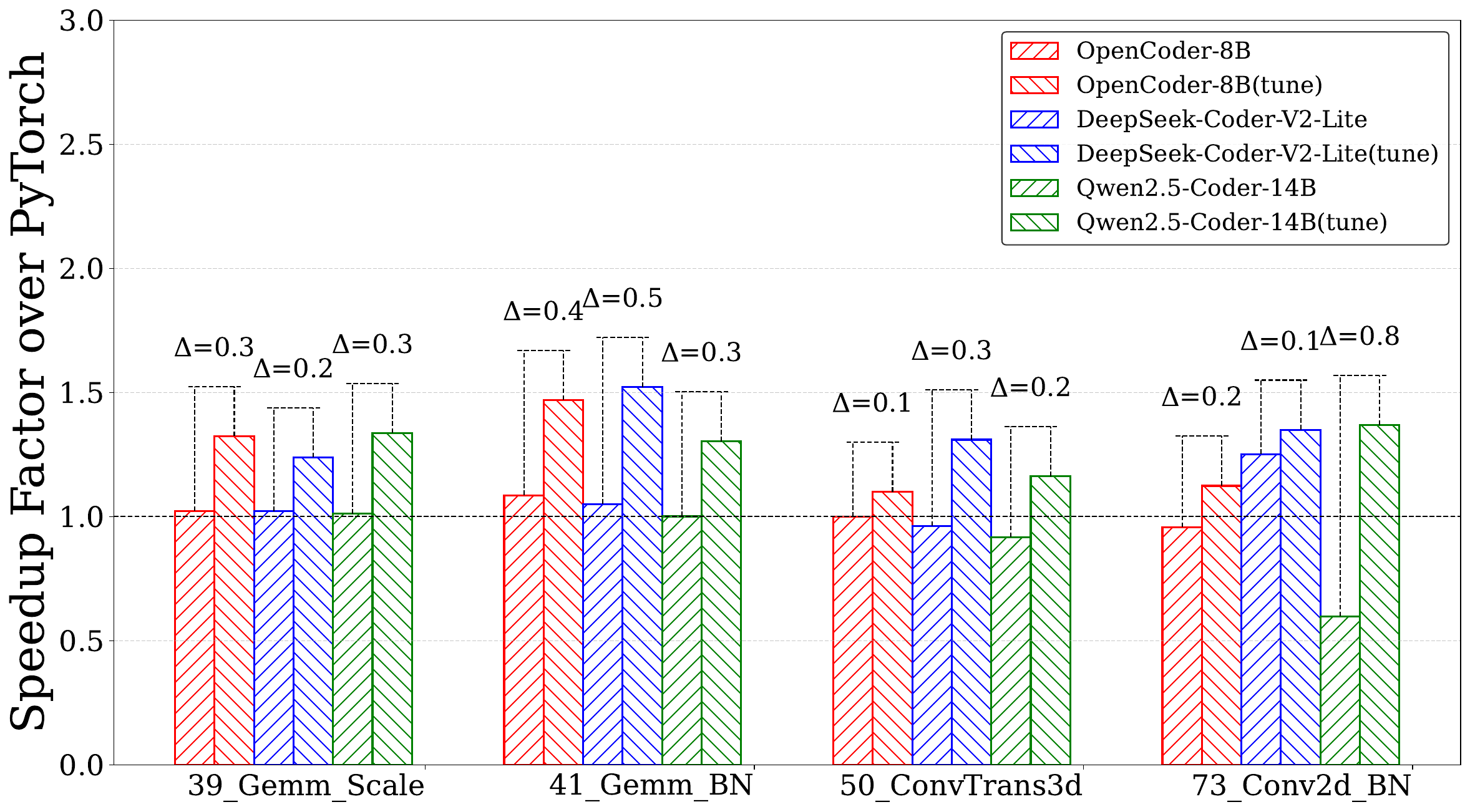}
        \caption*{(a) Level2 Performance Comparison}
        \label{fig:optim_level2_kernelbench}
    \end{minipage}
    \hfill
    \begin{minipage}{0.47\textwidth}
        \centering
        \includegraphics[width=\linewidth]{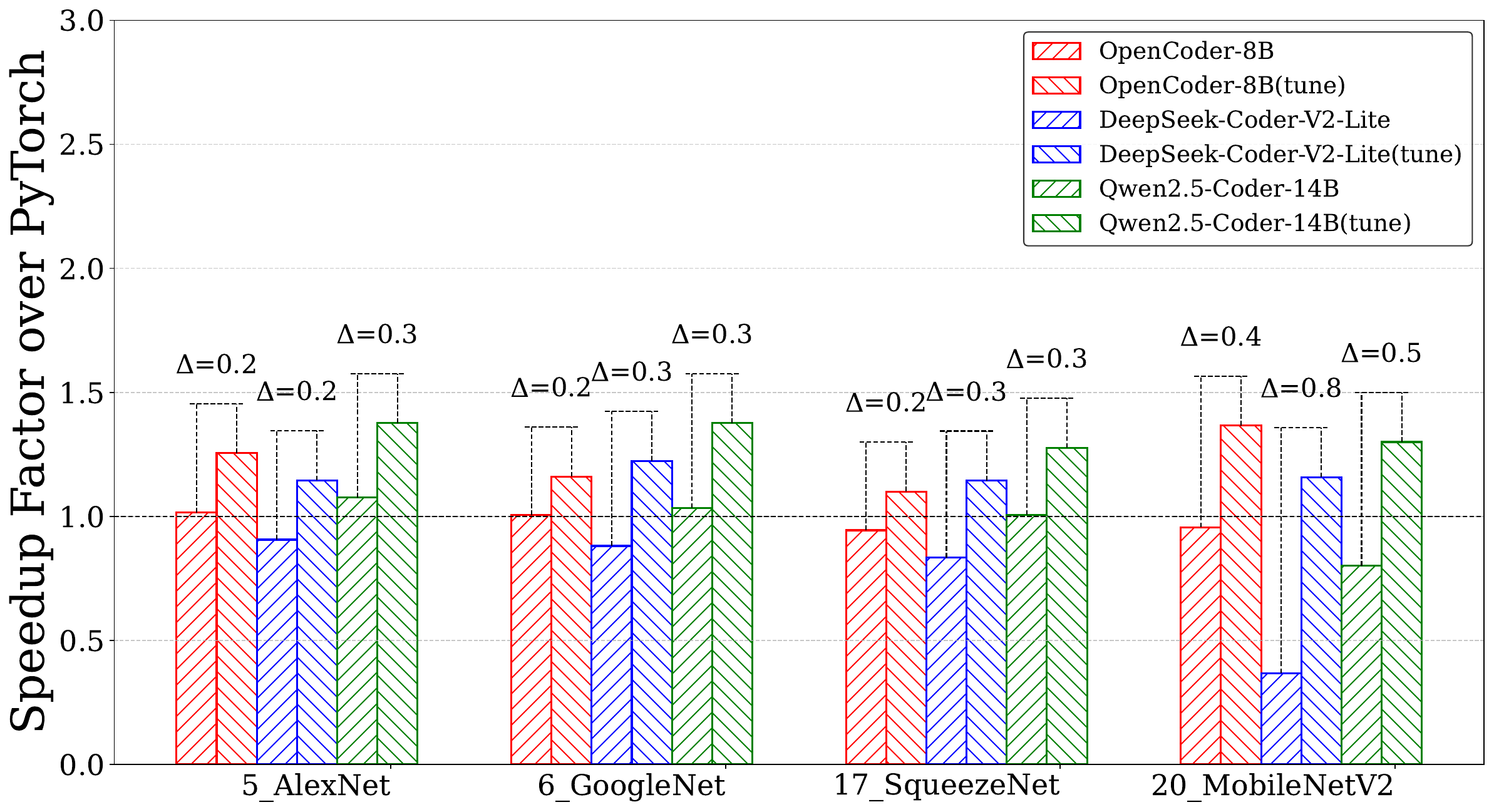}
        \caption*{(b) Level3 Performance Comparison}
        \label{fig:optim_level3_kernelbench}
    \end{minipage}
    \caption{Performance Comparison of Qwen2.5-Coder-14B, DeepSeek-Coder-V2-Lite, and OpenCoder-8B Models Before and After Fine-Tuning On Level2 and Level3}
    \label{fig:optim_kernelbench}
\end{figure*}

\begin{figure}
    \centering
    \includegraphics[width=0.97\linewidth]{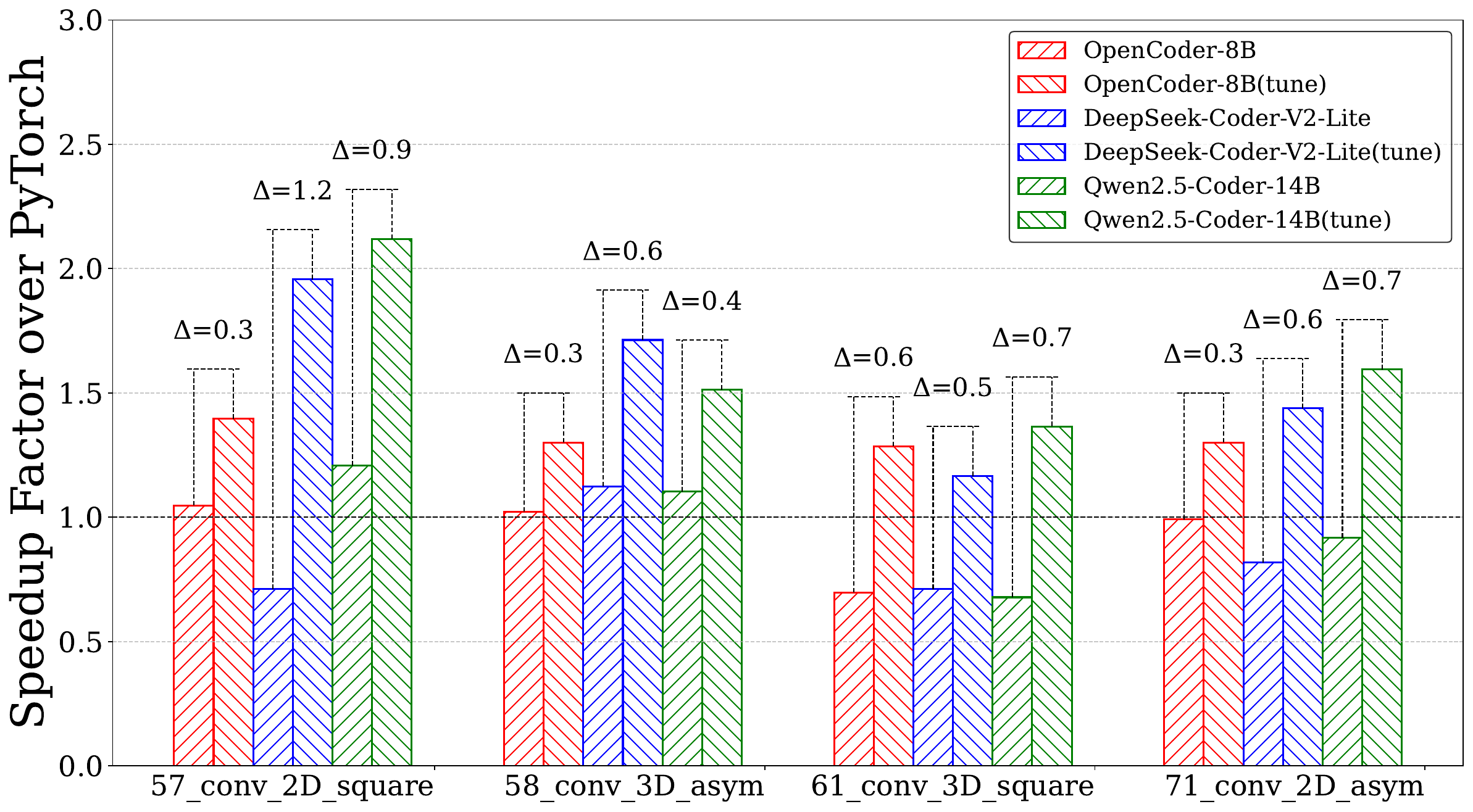} % 设置宽度并指定图片路径
    \caption{Performance Comparison of Qwen2.5-Coder-14B, DeepSeek-Coder-V2-Lite, and OpenCoder-8B Models Before and After Fine-Tuning On Level1} % 标题
    \label{fig:optim_level1_kernelbench} % 标签（用于交叉引用）
\end{figure}

The performance optimization results of $KernelBench_c$ are presented in Figures \ref{fig:optim_kernelbench} and \ref{fig:optim_level1_kernelbench}. Across all three levels of experiments, Qwen2.5-Coder-14B achieves the best performance after fine-tuning, with a 0.9x speedup in Level 1's 2D square convolutions, a 0.8x speedup in Level 2's Conv2d BN operation, and a 0.5x speedup in Level 3's MobileNetV2 inference. These results consistently outperform other models, highlighting the model's ability to optimize computation intensive tasks by leveraging operator-specific optimizations. The large performance gains, particularly in Level 1 operations like 2D/3D convolutions, can be attributed to the intrinsic parallelism of these operators, which Qwen2.5-Coder-14B effectively exploits through fine-tuning to adapt to specialized hardware instructions and efficient memory access patterns. In Level 2, the moderate improvements for Gemm-Scale and Conv-Trans are due to their higher reliance on matrix multiplications and intermediate data manipulations, where fine-tuning enhances kernel fusion and execution efficiency. At Level 3, while the overall improvement is less pronounced, the gains in MobileNetV2 inference demonstrate the ability of fine-tuning to adapt the model to diverse layer structures and operations. The other two models also demonstrate varying degrees of performance improvement after fine-tuning.

\begin{table}
\caption{Performance comparison on different operator types in Level1 before and after $HPCTransEval$ fine-tuning.}
\centering
\begin{tabular}{c|c|ccc}
\toprule
\multirow{2}{*}{Operator Type} & \multirow{2}{*}{Tuning} & \multicolumn{3}{c}{Model}  \\ 
\cmidrule(lr){3-5}
    &   &  \multicolumn{1}{c|}{$\mathcal{M}_1$} & \multicolumn{1}{c|}{$\mathcal{M}_2$} & \multicolumn{1}{c}{$\mathcal{M}_3$}   \\ 
\midrule
\multirow{2}{*}{Level 1 $[Elementwise]$}      
                        & $Before$           & 0.94     & 0.97     & 0.87   \\
                        & $After$               & 1.07     & 1.12     & 0.95  \\

\midrule
\multirow{2}{*}{Level 1 $[Reduction]$}  
                        & $Before$               & 0.84     & 0.93     & 0.86  \\
                        & $After$               & 1.04     & 1.58     & 1.34  \\
\midrule
\multirow{2}{*}{Level 1 $[Layout Transform]$}  
                        & $Before$               & 0.74     & 0.76     & 0.54  \\
                        & $After$               & 1.53     & 2.38     & 1.76  \\
\midrule
\multirow{2}{*}{Level 1  $[Logic Intensive]$}  
                        & $Before$               & 0.89     & 1.16     & 1.04  \\
                        & $After$               & 2.06     & 1.75     & 1.64  \\
\midrule
\multirow{2}{*}{Level 1 $[Compute Intensive]$}  
                        & $Before$               & 0.72     & 0.76     & 0.83  \\
                        & $After$               & 3.19     & 2.97     & 2.52  \\
\bottomrule
\end{tabular}
\label{tab:fine_tune_comparison_datasets}
\end{table}

As shown in Table \ref{tab:fine_tune_comparison_datasets}, performance improvements after fine-tuning on $HPCTransEval$ demonstrate varying optimization potential between operator types. Elementwise operators, with their modest improvement of up to 15.5\%, reaffirm their computational simplicity and limited scope for enhancement. The inherent challenges of reduction operators, such as managing efficient memory access and achieving optimal parallel execution, are effectively addressed, resulting in performance gains reaching 69.9\%. The story turns even more dramatic with Layout Transform operators, where performance surges by as much as 225.9\%, exposing the hidden inefficiencies in default memory layouts that fine-tuning so effectively resolves. Logic Intensive operators add their own twist, achieving up to 131.5\% improvement through refined control flow and minimized branching penalties. The Compute Intensive operators demonstrate significant improvements, exceeding 340\%, highlighting the substantial impact of fine-tuning in enhancing computational efficiency. The performance disparities across operator types suggest targeted adjustments in dataset design for further fine-tuning. The fine-tuning datasets should maintain a balanced representation of operator types while skewing towards those with the most substantial optimization potential.

\subsection{Detailed Case Study}

This section presents two detailed case studies demonstrating that fine-tuned Models within our framework can translate CUDA kernels for CPU execution while simultaneously optimizing the generated code for CPU hardware characteristics. We analyze two representative examples: (1) The translation of a grouped kernel comprising [Matmul, Dropout, Mean, Softmax] by Qwen2.5-Coder-14B, comparing outputs before and after fine-tuning, and (2) The translation of the VGG-16 convolutional neural network by DeepSeek-Coder-V2-Lite, again comparing outputs pre- and post-fine-tuning. These examples illustrate how fine-tuning enables the models to generate CPU code that better leverages platform-specific features, leading to improved performance.

\begin{lstlisting}[label=code:group-qwen, language=C++, caption={Code for Matlmul+Dropout+Mean+Softmax before Qwen2.5-Coder-14B fine-tuning.}, breaklines=true]
void forward_cpu(
    // ...
    #pragma omp parallel for
    for (...) {
        for (...) {
            // ...
            unsigned long long offset = i * out_features + j;
            std::mt19937 gen(seed + offset);
            std::uniform_real_distribution<float> dis(0.0f, 1.0f);
            float rand_val = dis(gen);
            // ...
        }
    }
}
\end{lstlisting}

The first case involves the fused [Matmul, Dropout, Mean, Softmax] kernel. Code \ref{code:group-qwen} displays the C++ code generated by the base Qwen2.5-Coder-14B model. A significant performance bottleneck arises from the repeated instantiation of std::mt19937 pseudo-random number generators (PRNGs) within the OpenMP parallel loop (\#pragma omp parallel for). Each instantiation incurs substantial overhead due to the Mersenne Twister algorithm's initialization process, which involves allocating and seeding a relatively large internal state (624 32-bit integers). Instantiating this within the parallel loop leads to high computational cost and potential memory contention.

In contrast, the code generated by the fine-tuned Qwen model (Code \ref{code:group-qwen-lora}) replaces the `std::mt19937` generator with a lightweight hash-based PRNG. This alternative relies on efficient bitwise operations and modular arithmetic, eliminating the significant state initialization and maintenance overhead associated with the Mersenne Twister. This optimization reduces register pressure, minimizes thread contention during PRNG initialization, and potentially enhances instruction-level parallelism. While the statistical quality of the hash-based PRNG might be lower than that of `std::mt19937`, it provides sufficient randomness for typical dropout layer implementations in neural networks, striking an effective balance between computational efficiency and the stochastic requirements of the application domain.

\begin{lstlisting}[label=code:group-qwen-lora, language=C++, caption={Code for Matlmul+Dropout+Mean+Softmax after Qwen2.5-Coder-14B fine-tuning.}, breaklines=true]
void forward_cpu(
    // ...
#pragma omp parallel for
  for (...) {
    for (...) {
        // ...
        unsigned long long offset = i * out_features + j;
        unsigned long long hash = offset + seed;
        hash ^= hash >> 17;
        hash ^= hash << 31;
        hash ^= hash >> 8;
        float rand_val = static_cast<float>(hash % 1000) / 1000.0f;
        // ...
      }
    }
  }
}
\end{lstlisting}

The second case study examines the translation of VGG-16 for CPU execution. The code generated by the base DeepSeek-Coder-V2-Lite model (Code \ref{code:vgg16-base}) reveals architectural mismatches stemming from mimicking CUDA programming patterns. Specifically, it employs techniques analogous to GPU shared memory (using std::vector potentially per thread) and complex tiling/indexing schemes. While effective on GPUs with explicitly managed local memories and massive parallelism, these patterns are often detrimental on CPUs, which rely heavily on hierarchical caches and hardware prefetching. The base model's approach disrupts data locality (both spatial and temporal), leading to poor cache utilization, increased cache misses, and memory latency penalties.

Conversely, the fine-tuned DeepSeek model generates code (Code \ref{code:vgg16-lora}) that adopts a more conventional, cache-friendly nested loop structure for the convolution operation. This structure aligns better with the sequential processing nature and memory hierarchy of CPUs. It promotes regular memory access patterns, enhancing data locality and maximizing the effectiveness of the CPU cache system. Furthermore, this straightforward structure is more amenable to compiler optimizations, such as automatic SIMD vectorization, further boosting performance. Both examples underscore the capability of the fine-tuned models to generate not just functionally correct translations, but performance-optimized code adapted to the specific architectural constraints of the target CPU platform.

\begin{lstlisting}[label=code:vgg16-base, language=C++, caption={Code for VGG-16 before DeepSeek-Coder-V2-Lite fine-tuning.}, breaklines=true]
void conv2d_atomic_kernel_cpu(...) {
    int tile_dim = tile_size * tile_size;
    std::vector<float> shared_mem(omp_get_max_threads() * tile_dim);

    #pragma omp parallel
    {
        int thread_id = omp_get_thread_num();
        int num_threads = omp_get_num_threads();
        int tile_start = thread_id * (H * W) / num_threads;
        int tile_end = (thread_id + 1) * (H * W) / num_threads;
        for (int tile_idx = tile_start; tile_idx < tile_end; tile_idx += H * W) {
        int h = tile_idx / W % H;
        int w = tile_idx % W;
        float sum = 0.0f;
        for (int c = 0; c < C; ++c) {
        for (int kh = 0; kh < 3; ++kh) {
        for (int kw = 0; kw < 3; ++kw) {
        int ih = h - P + kh;
        int iw = w - P + kw;
        if (ih >= 0 && ih < H && iw >= 0 && iw < W) {
            // ...
        }
        }
        }
        }
        sum += bias[K * (tile_idx / (H * W)) + (tile_idx / (H * W))];
        output[tile_idx] = sum;
        }
    }
}
\end{lstlisting}

\begin{lstlisting}[label=code:vgg16-lora, language=C++, caption={Code for VGG-16 after DeepSeek-Coder-V2-Lite fine-tuning.}, breaklines=true]
void conv2d_atomic_kernel_cpu(const float* input, const float* weight, const float* bias, float* output, int N, int C, int H, int W, int K, int P, int stride) {
    #pragma omp parallel for
    for (int n = 0; n < N; ++n) {
    for (int k = 0; k < K; ++k) {
    for (int h = 0; h < H; ++h) {
    for (int w = 0; w < W; ++w) {
    float sum = bias[k];
    for (int c = 0; c < C; ++c) {
    for (int kh = 0; kh < 3; ++kh) {
    for (int kw = 0; kw < 3; ++kw) {
        // ...
    }
    }
    }
    output[n * K * H * W + k * H * W + h * W + w] = sum;
    }
    }
    }
    }
}
\end{lstlisting}

\subsection{Scalability Analysis}

\begin{figure}
    \centering
    \begin{minipage}{0.47\textwidth}
        \centering
        \includegraphics[width=\linewidth]{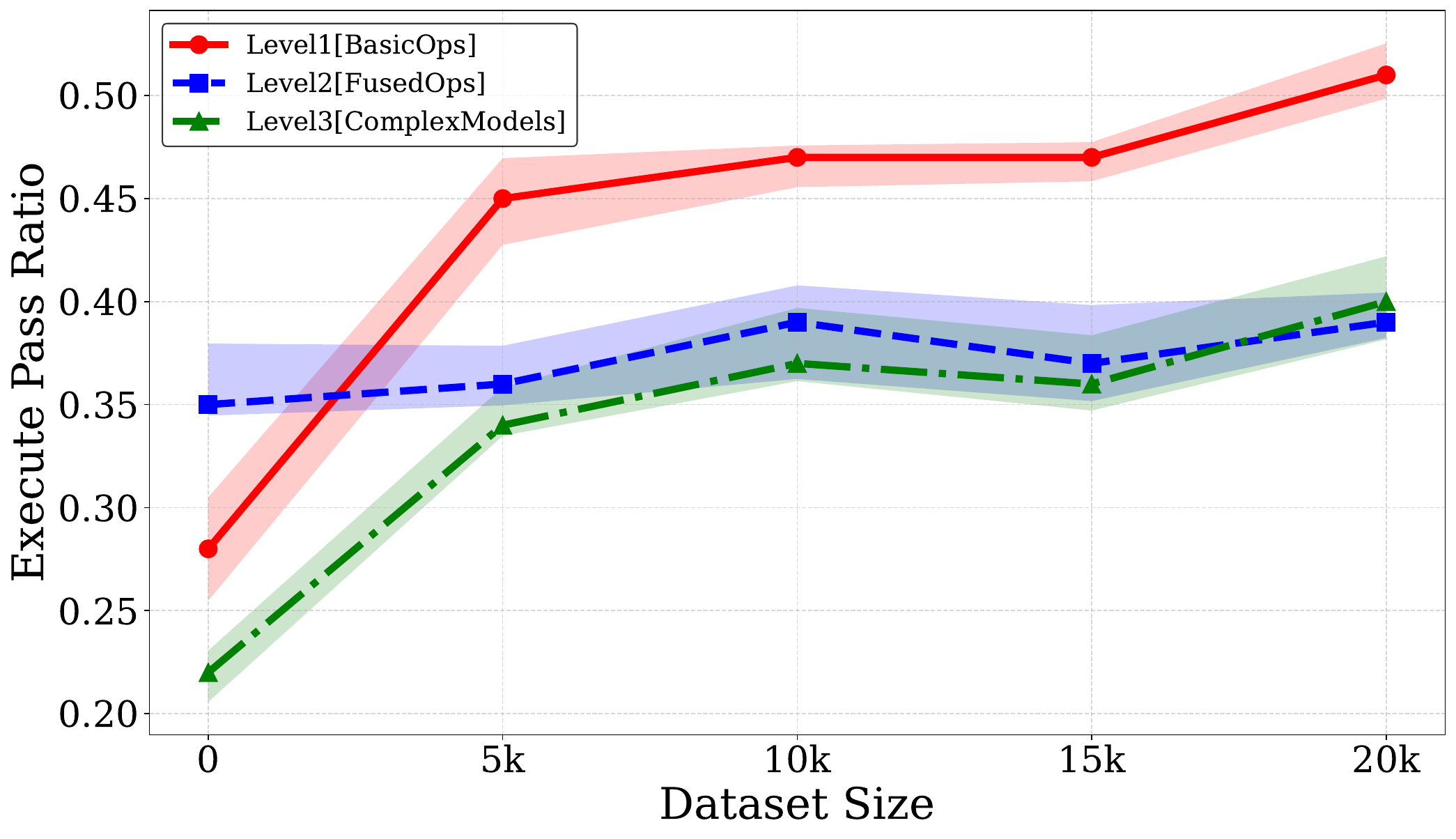}
        \caption*{(a) Execute Pass Ratio under different training set proportions.}
        \label{fig:homo_a}
    \end{minipage}
    \hfill
    \begin{minipage}{0.47\textwidth}
        \centering
        \includegraphics[width=\linewidth]{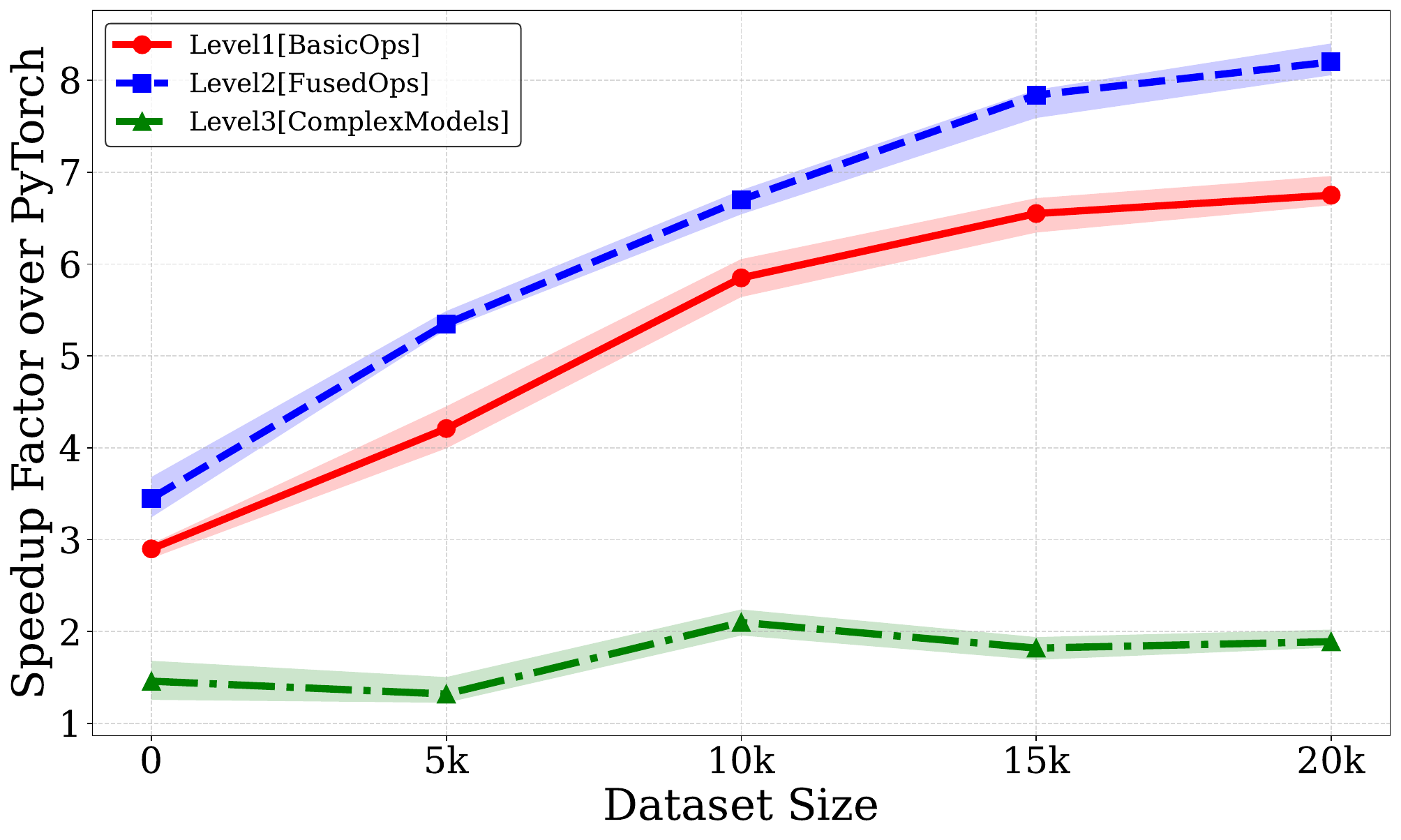}
        \caption*{(b) Speedup Factor over Pytorch under different training set proportions.}
        \label{fig:homo_b}
    \end{minipage}
    \caption{Qwen2.5-Coder-14B execute pass and speedup factor under different training set proportions.}
    \label{fig:homo}
\label{fig:scalability_analysis}
\end{figure}

As shown in Figure \ref{fig:scalability_analysis}, we further analyze the impact of dataset size on the fine-tuning results on $KernelBench_c$. It can be observed that when the model is trained with a dataset size of 5k, the operator execution success rate reaches a relatively high level. As the dataset size increases, the improvement in operator correctness becomes marginal. Additionally, since the Level 1 translation task is relatively simpler, increasing the dataset size yields more significant benefits. Regarding acceleration, both Level 1 and Level 2 code performance continuously improve with the increasing dataset size and stabilize around 20k. On the other hand, the performance of Level 3 code does not change significantly, likely because the operator at the model level still relies on the underlying PyTorch interfaces during actual execution. Injecting domain knowledge into LLMs mainly enhances their ability to apply certain optimization techniques, without affecting the choice of interface calls. In general, increasing the dataset size consistently has a positive impact on the performance of the generated code.

\section{Related Work}

\paragraph{CUDA Ecosystem Compatibility}

Several methods have been proposed to migrate the CUDA ecosystem to CPUs or other platforms. The most direct way is to re-engineer the target application, such as the Intel oneDNN library \cite{onednn}, which can lead to huge development costs. For GPGPU hardware like AMD, they have developed the ROCm \cite{ROCm} programming interface, which is very similar to Nvidia CUDA. Additionally, they provide HIP \cite{hip}, which is a C++ runtime API and kernel language that allows developers to create portable applications for both AMD and NVIDIA GPUs from a single source code. To seamlessly support CUDA, HIPIFY \cite{HIPIFY} can translate CUDA source code to HIP source code via compiler frontend technology, such as AST transformation. 

% It is not possible to simply achieve CUDA software compatibility on CPUs by replacing APIs due to differences both in hardware architecture and programming model. From the software perspective, even though users program in C language, CUDA adopts the single program multiple data (SPMD) programming model, while the fork-join programming model is used on CPUs. There exist many approaches for performance portability such as language extensions (e.g., OpenCL \cite{opencl}, SYCL \cite{sycl}), domain-specific languages (e.g., Halide \cite{ragan2013halide}, TVM \cite{chen2018tvm}), parallel programming frameworks (e.g., Kokkos \cite{edwards2014kokkos}). However, all these approaches still require some rewriting via specific library APIs or DSLs, which is infeasible for large and complex applications or those that are actively being updated. The most advanced and automatic method is the Polygeist \cite{moses2023poly} compiler which can transpile CUDA code to CPU C code via encoding high-level parallel constructs in its intermediate representations (IRs). However, this approach is limited by the completeness of the compiler implementation, and the design of its front-end and intermediate representations also limits the range of CUDA workloads it can handle.

It is not possible to simply achieve CUDA software compatibility on CPUs by replacing APIs due to differences both in hardware architecture and programming model. From the software perspective, even though users program in C language, CUDA adopts the single program multiple data (SPMD) programming model, while the fork-join programming model is used on CPUs. There exist many approaches for performance portability such as language extensions (e.g., OpenCL \cite{opencl}, SYCL \cite{sycl}), domain-specific languages (e.g., Halide \cite{ragan2013halide}, TVM \cite{chen2018tvm}), and parallel programming frameworks (e.g., Kokkos). However, all these approaches still require some rewriting via specific library APIs or DSLs, which is infeasible for large and complex applications or those that are actively being updated. The most advanced and automatic method is the Polygeist compiler which can transpile CUDA code to CPU C code via encoding high-level parallel constructs in its intermediate representations (IRs). However, this approach is limited by the completeness of the compiler implementation, and the design of its front-end and intermediate representations also limits the range of CUDA workloads it can handle.

% \paragraph{Benchmark for Code Tasks}
% HumanEval \cite{zheng2023codegeex} is a benchmark to measure functional correctness for synthesizing programs from docstrings. It consists of 164 original programming problems, assessing language comprehension, algorithms, and simple mathematics, with some comparable to simple software interview questions.

% CodeXGLUE \cite{lu2021codexglue} is a benchmark dataset and an open challenge for code intelligence. It includes a collection of code intelligence tasks and a platform for model evaluation and comparison. CodeXGLUE stands for General Language Understanding Evaluation benchmark for CODE. It includes 14 datasets for 10 diversified code intelligence tasks covering the following scenarios: code-code, code-text, text-code, and text-text. 

% xCodeEval \cite{khan2023xcodeeval} is one of the largest executable multilingual multitask benchmarks consisting of 17 programming languages with execution-level parallelism. It features a total of seven tasks involving code understanding, generation, translation, and retrieval, and it employs an execution-based evaluation instead of traditional lexical approaches. It also provides a test-case-based multilingual code execution engine, ExecEval that supports all the programming languages in xCodeEval.

\section{Conclusion}

In response to the challenges of translating CUDA-based GPU code to other platforms for high-performance computing, our proposed framework leverages TVM's operator optimization and graph-based data augmentation to generate high-quality CUDA and corresponding platform code pairs, addressing the limitations of existing LLMs caused by insufficient training datasets. The introduction of HPCTransEval provides a robust benchmark for evaluating LLM performance in CUDA transpilation. Experimental results demonstrate significant improvements in transpilation quality, underscoring the potential of integrating specialized optimization techniques and domain-specific knowledge into LLMs. This approach not only enhances performance portability across diverse hardware platforms, but also paves the way for more efficient and scalable solutions in parallel computing, aligning with the growing computational demands of modern deep learning models.


\begin{thebibliography}{10}

\bibitem{amazon}
Amazon codewhisperer home page.
\newblock \url{https://aws.amazon.com/pm/codewhisperer/}, Amazon.2023.

\bibitem{hip}
\textsc{AMD HIP}.
\newblock \url{https://github.com/ROCm/HIP}, AMD.2016.

\bibitem{ROCm}
\textsc{AMD ROC}m.
\newblock \url{https://www.amd.com/en/products/software/rocm.html}, AMD.2016.

\bibitem{CodeGeeX}
Codegeex home page.
\newblock \url{https://codegeex.cn/en-US/}, CodeGeeX.2022.

\bibitem{copilot}
Github copilot home page.
\newblock \url{https://github.com/features/copilot}, Github.2021.

\bibitem{HIPIFY}
\textsc{HIPIFY} github home page.
\newblock \url{https://github.com/ROCm/HIPIFY}, HIPIFY.2016.

\bibitem{onednn}
Intel. 2022. \textsc{O}ne\textsc{API} \textsc{D}eep \textsc{N}eural \textsc{N}etwork \textsc{L}ibrary (\textsc{O}ne\textsc{DNN}).
\newblock \url{https://github.com/oneapi-src/oneDNN}, Intel.2022.

\bibitem{orcjit}
\textsc{LLVM ORC JIT} documentation.
\newblock \url{https://llvm.org/docs/ORCv2.html}, LLVM.2018.

\bibitem{sycl}
\textsc{SYCL} home page.
\newblock \url{https://www.khronos.org/sycl/}, SYCL.2014.

\bibitem{ahmed2024automatic}
{\sc Ahmed, T., Pai, K.~S., Devanbu, P., and Barr, E.}
\newblock Automatic semantic augmentation of language model prompts (for code summarization).
\newblock In {\em Proceedings of the IEEE/ACM 46th International Conference on Software Engineering\/} (2024), pp.~1--13.

\bibitem{claude35}
{\sc Anthropic}.
\newblock claude 3.5 home page.
\newblock \url{https://www.anthropic.com/claude}.

\bibitem{chen2021evaluating}
{\sc Chen, M., Tworek, J., Jun, H., Yuan, Q., Pinto, H. P. D.~O., Kaplan, J., Edwards, H., Burda, Y., Joseph, N., Brockman, G., et~al.}
\newblock Evaluating large language models trained on code.
\newblock {\em arXiv preprint arXiv:2107.03374\/} (2021).

\bibitem{chen2018tvm}
{\sc Chen, T., Moreau, T., Jiang, Z., Zheng, L., Yan, E., Shen, H., Cowan, M., Wang, L., Hu, Y., Ceze, L., et~al.}
\newblock \textsc{TVM}: An automated end-to-end optimizing compiler for deep learning.
\newblock In {\em 13th USENIX Symposium on Operating Systems Design and Implementation (OSDI 18)\/} (2018), pp.~578--594.

\bibitem{chen2018tree}
{\sc Chen, X., Liu, C., and Song, D.}
\newblock Tree-to-tree neural networks for program translation.
\newblock {\em Advances in neural information processing systems 31\/} (2018).

\bibitem{chen2024large}
{\sc Chen, Y., Wu, J., Ling, X., Li, C., Rui, Z., Luo, T., and Wu, Y.}
\newblock When large language models confront repository-level automatic program repair: How well they done?
\newblock In {\em Proceedings of the 2024 IEEE/ACM 46th International Conference on Software Engineering: Companion Proceedings\/} (2024), pp.~459--471.

\bibitem{edwards2014kokkos}
{\sc Edwards, H.~C., Trott, C.~R., and Sunderland, D.}
\newblock Kokkos: Enabling manycore performance portability through polymorphic memory access patterns.
\newblock {\em Journal of parallel and distributed computing 74}, 12 (2014), 3202--3216.

\bibitem{guo2023longcoder}
{\sc Guo, D., Xu, C., Duan, N., Yin, J., and McAuley, J.}
\newblock Longcoder: A long-range pre-trained language model for code completion.
\newblock In {\em International Conference on Machine Learning\/} (2023), PMLR, pp.~12098--12107.

\bibitem{hu2021lora}
{\sc Hu, E.~J., Shen, Y., Wallis, P., Allen-Zhu, Z., Li, Y., Wang, S., Wang, L., and Chen, W.}
\newblock Lora: Low-rank adaptation of large language models.
\newblock {\em arXiv preprint arXiv:2106.09685\/} (2021).

\bibitem{huang2024opencoder}
{\sc Huang, S., Cheng, T., Liu, J.~K., Hao, J., Song, L., Xu, Y., Yang, J., Liu, J., Zhang, C., Chai, L., et~al.}
\newblock Opencoder: The open cookbook for top-tier code large language models.
\newblock {\em arXiv preprint arXiv:2411.04905\/} (2024).

\bibitem{hui2024qwen2}
{\sc Hui, B., Yang, J., Cui, Z., Yang, J., Liu, D., Zhang, L., Liu, T., Zhang, J., Yu, B., Lu, K., et~al.}
\newblock Qwen2. 5-coder technical report.
\newblock {\em arXiv preprint arXiv:2409.12186\/} (2024).

\bibitem{jiang2021cure}
{\sc Jiang, N., Lutellier, T., and Tan, L.}
\newblock Cure: Code-aware neural machine translation for automatic program repair.
\newblock In {\em 2021 IEEE/ACM 43rd International Conference on Software Engineering (ICSE)\/} (2021), IEEE, pp.~1161--1173.

\bibitem{jouppi2023tpu}
{\sc Jouppi, N., Kurian, G., Li, S., Ma, P., Nagarajan, R., Nai, L., Patil, N., Subramanian, S., Swing, A., Towles, B., et~al.}
\newblock Tpu v4: An optically reconfigurable supercomputer for machine learning with hardware support for embeddings.
\newblock In {\em Proceedings of the 50th Annual International Symposium on Computer Architecture\/} (2023), pp.~1--14.

\bibitem{kaplan2020scalinglaws}
{\sc Kaplan, J., McCandlish, S., Henighan, T., Brown, T.~B., Chess, B., Child, R., Gray, S., Radford, A., Wu, J., and Amodei, D.}
\newblock Scaling laws for neural language models.
\newblock {\em CoRR abs/2001.08361\/} (2020).

\bibitem{khan2023xcodeeval}
{\sc Khan, M. A.~M., Bari, M.~S., Do, X.~L., Wang, W., Parvez, M.~R., and Joty, S.}
\newblock xcodeeval: A large scale multilingual multitask benchmark for code understanding, generation, translation and retrieval.
\newblock {\em arXiv preprint arXiv:2303.03004\/} (2023).

\bibitem{kulal2019spoc}
{\sc Kulal, S., Pasupat, P., Chandra, K., Lee, M., Padon, O., Aiken, A., and Liang, P.~S.}
\newblock Spoc: Search-based pseudocode to code.
\newblock {\em Advances in Neural Information Processing Systems 32\/} (2019).

\bibitem{laso2024exploring}
{\sc Laso, R., Krupitza, D., and Hunold, S.}
\newblock Exploring scalability in c++ parallel stl implementations.
\newblock In {\em Proceedings of the 53rd International Conference on Parallel Processing\/} (2024), pp.~284--293.

\bibitem{loshchilov2017decoupled}
{\sc Loshchilov, I., and Hutter, F.}
\newblock Decoupled weight decay regularization.
\newblock {\em arXiv preprint arXiv:1711.05101\/} (2017).

\bibitem{lozhkov2024starcoder}
{\sc Lozhkov, A., Li, R., Allal, L.~B., Cassano, F., Lamy-Poirier, J., Tazi, N., Tang, A., Pykhtar, D., Liu, J., Wei, Y., et~al.}
\newblock Starcoder 2 and the stack v2: The next generation.
\newblock {\em arXiv preprint arXiv:2402.19173\/} (2024).

\bibitem{lu2021codexglue}
{\sc Lu, S., Guo, D., Ren, S., Huang, J., Svyatkovskiy, A., Blanco, A., Clement, C., Drain, D., Jiang, D., Tang, D., et~al.}
\newblock Codexglue: A machine learning benchmark dataset for code understanding and generation.
\newblock {\em arXiv preprint arXiv:2102.04664\/} (2021).

\bibitem{moses2023poly}
{\sc Moses, W.~S., Ivanov, I.~R., Domke, J., Endo, T., Doerfert, J., and Zinenko, O.}
\newblock High-performance gpu-to-cpu transpilation and optimization via high-level parallel constructs.
\newblock In {\em Proceedings of the 28th ACM SIGPLAN Annual Symposium on Principles and Practice of Parallel Programming\/} (2023), pp.~119--134.

\bibitem{ouyang2025kernelbench}
{\sc Ouyang, A., Guo, S., Arora, S., Zhang, A.~L., Hu, W., R{\'e}, C., and Mirhoseini, A.}
\newblock Kernelbench: Can llms write efficient gpu kernels?
\newblock {\em arXiv preprint arXiv:2502.10517\/} (2025).

\bibitem{pan2024lost}
{\sc Pan, R., Ibrahimzada, A.~R., Krishna, R., Sankar, D., Wassi, L.~P., Merler, M., Sobolev, B., Pavuluri, R., Sinha, S., and Jabbarvand, R.}
\newblock Lost in translation: A study of bugs introduced by large language models while translating code.
\newblock In {\em Proceedings of the IEEE/ACM 46th International Conference on Software Engineering\/} (2024), pp.~1--13.

\bibitem{ragan2013halide}
{\sc Ragan-Kelley, J., Barnes, C., Adams, A., Paris, S., Durand, F., and Amarasinghe, S.}
\newblock Halide: a language and compiler for optimizing parallelism, locality, and recomputation in image processing pipelines.
\newblock {\em Acm Sigplan Notices 48}, 6 (2013), 519--530.

\bibitem{ren2020codebleu}
{\sc Ren, S., Guo, D., Lu, S., Zhou, L., Liu, S., Tang, D., Sundaresan, N., Zhou, M., Blanco, A., and Ma, S.}
\newblock Code\textsc{BLEU}: a method for automatic evaluation of code synthesis.
\newblock {\em arXiv preprint arXiv:2009.10297\/} (2020).

\bibitem{shazeer2017outrageously}
{\sc Shazeer, N., Mirhoseini, A., Maziarz, K., Davis, A., Le, Q., Hinton, G., and Dean, J.}
\newblock Outrageously large neural networks: The sparsely-gated mixture-of-experts layer.
\newblock {\em arXiv preprint arXiv:1701.06538\/} (2017).

\bibitem{opencl}
{\sc Stone, J.~E., Gohara, D., and Shi, G.}
\newblock Open\textsc{CL}: A parallel programming standard for heterogeneous computing systems.
\newblock {\em Computing in Science \& Engineering 12}, 3 (2010), 66--73.

\bibitem{vaswani2017attention}
{\sc Vaswani, A., Shazeer, N., Parmar, N., Uszkoreit, J., Jones, L., Gomez, A.~N., Kaiser, {\L}., and Polosukhin, I.}
\newblock Attention is all you need.
\newblock {\em Advances in neural information processing systems 30\/} (2017).

\bibitem{wolf2020transformers}
{\sc Wolf, T., Debut, L., Sanh, V., Chaumond, J., Delangue, C., Moi, A., Cistac, P., Rault, T., Louf, R., Funtowicz, M., et~al.}
\newblock Transformers: State-of-the-art natural language processing.
\newblock In {\em Proceedings of the 2020 conference on empirical methods in natural language processing: system demonstrations\/} (2020), pp.~38--45.

\bibitem{yu2024enhancing}
{\sc Yu, C., Royuela, S., and Qui{\~n}ones, E.}
\newblock Enhancing heterogeneous computing through openmp and gpu graph.
\newblock In {\em Proceedings of the 53rd International Conference on Parallel Processing\/} (2024), pp.~534--543.

\bibitem{yu2024codereval}
{\sc Yu, H., Shen, B., Ran, D., Zhang, J., Zhang, Q., Ma, Y., Liang, G., Li, Y., Wang, Q., and Xie, T.}
\newblock Codereval: A benchmark of pragmatic code generation with generative pre-trained models.
\newblock In {\em Proceedings of the 46th IEEE/ACM International Conference on Software Engineering\/} (2024), pp.~1--12.

\bibitem{zhang2021cpm}
{\sc Zhang, Z., Han, X., Zhou, H., Ke, P., Gu, Y., Ye, D., Qin, Y., Su, Y., Ji, H., Guan, J., et~al.}
\newblock Cpm: A large-scale generative chinese pre-trained language model.
\newblock {\em AI Open 2\/} (2021), 93--99.

\bibitem{zheng2020ansor}
{\sc Zheng, L., Jia, C., Sun, M., Wu, Z., Yu, C.~H., Haj-Ali, A., Wang, Y., Yang, J., Zhuo, D., Sen, K., et~al.}
\newblock Ansor: Generating high-performance tensor programs for deep learning.
\newblock In {\em 14th USENIX symposium on operating systems design and implementation (OSDI 20)\/} (2020), pp.~863--879.

\bibitem{zheng2023codegeex}
{\sc Zheng, Q., Xia, X., Zou, X., Dong, Y., Wang, S., Xue, Y., Shen, L., Wang, Z., Wang, A., Li, Y., et~al.}
\newblock Codegeex: A pre-trained model for code generation with multilingual benchmarking on humaneval-x.
\newblock In {\em Proceedings of the 29th ACM SIGKDD Conference on Knowledge Discovery and Data Mining\/} (2023), pp.~5673--5684.

\bibitem{zhou2022large}
{\sc Zhou, Y., Muresanu, A.~I., Han, Z., Paster, K., Pitis, S., Chan, H., and Ba, J.}
\newblock Large language models are human-level prompt engineers.
\newblock {\em arXiv preprint arXiv:2211.01910\/} (2022).

\bibitem{zhu2024deepseek}
{\sc Zhu, Q., Guo, D., Shao, Z., Yang, D., Wang, P., Xu, R., Wu, Y., Li, Y., Gao, H., Ma, S., et~al.}
\newblock Deepseek-coder-v2: Breaking the barrier of closed-source models in code intelligence.
\newblock {\em arXiv preprint arXiv:2406.11931\/} (2024).

\end{thebibliography}
\end{document}